\documentclass[conference]{IEEEtran}
\IEEEoverridecommandlockouts
\usepackage{cite}
\usepackage{amsmath,amssymb,amsfonts}
\usepackage{algorithmic}
\usepackage{graphicx}
\usepackage{textcomp}
\usepackage{xcolor}
\def\BibTeX{{\rm B\kern-.05em{\sc i\kern-.025em b}\kern-.08em
    T\kern-.1667em\lower.7ex\hbox{E}\kern-.125emX}}
\begin{document}

\title{High-Speed (7,2) Compressor Using A Fast Carry-Generation Logic based on Sorting Network}

\author{\IEEEauthorblockN{Wenbo Guo}
\IEEEauthorblockA{\textit{School of Integrated Circuits} \\
\textit{Tsinghua University}\\
Beijing, China \\
gwb.eip@foxmail.com}
}

\maketitle

\begin{abstract}
Fast binary compressors are the main components of many basic digital calculation units.
In this paper, a high-speed (7,2) compressor with a fast carry-generation logic is proposed. The carry-generation logic is based on the sorting network, and it can generate a carry bit within 2 logical stages other than 3 stages as in previous school book full adders. Collaborating with the adjusted full adder logic, the proposed (7,2) compressor achieves using only 11 basic logical stages.
Testing this new design in a binary arry with 7 rows and 8 columns, and the result shows that this design have higher proformance than previous designs. This method is suitable for high proformance cases in multiplication design or other cryptography hardware blocks.
\end{abstract}


\section{Introduction}
Multiplication is a very common operation in digital devices. And the performance of multiplication is the bottleneck of DSP. Fast multiplier consists of 3 parts:  partial product generation, partial product reduction, and vector merge addition. Wallace Tree \cite{b0} \cite{b00} is proposed to parallelly compress the partial product with full adders, and the full adders are now known as the (3,2) compressors. Thereafter, various methods are proposed to construct a more efficient compressor to further speed-up the reduction of partical products. Larger compressors have been widely used, such as (4,2), (5,2), (7,2) \cite{b01} \cite{b02} \cite{b03} compressors, but this reduction step still spend the most time in a multiplication operation. Besides, in many cryptography hardware blocks, such as modular multiplication, also require high-speed compressors to speed them up.

(7,2) compressor is proved to be a hign efficient method \cite{b1}, and many papers have discussed on this. Booth \cite{b1} and \cite{b2} are using special mechods to reduce the number of basic logical stages. The method in \cite{b2} reduced the logical layer to 12, but still using too many XOR gate which is slow and hard to optimize in logical level. The method in \cite{b1} also implemente a 12-logical-layer design with the help of (7,3) counter in \cite{b3}. But this design did not optimize for it.

Further reduce the basic logical layers of (7,2) compressors is difficult. However, it is still possible by using a special carry-generation logical. The contributions of this paper are listed below:

\begin{itemize}
  \item[1.] We proposed a carry-generation logic that generates a carry bit within 2 basic logical stages while traditional full adders require 3 basic logical stages.
  
  \item[2.] The adjusted full adder is introduced. It is designed to collaborate with the carry-generation unit.

  \item[3.] We propose the new (7,2) compressor that only consumes 11 basic ligical stages. According to the synthesis result, which is synthsised by Synopsys DC, our design has less time delay.

\end{itemize}

This paper will present our design firstly, and then present the comparison between our design and \cite{b1} or \cite{b2}.

\section{Method and implementation}

\subsection{Sorting Network of 1-bit Numbers}

\begin{figure}[htbp]
\centerline{\includegraphics[scale=0.4]{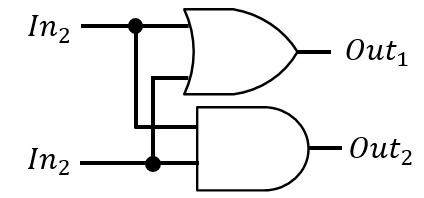}}
\caption{Sorter of two 1-bit numbers.}
\label{fig:sorter}
\end{figure}

For two 1-bit numbers, sorting is a simple operation. The circuit in Fig. \ref{fig:sorter} can easily sort the two input 1-bit numbers. $Out_1$ is always the larger one and $Out_2$ is always the minor one. This circuit only consumes an AND gate and an OR gate. 

\begin{figure}[htbp]
\centerline{\includegraphics[scale=0.30]{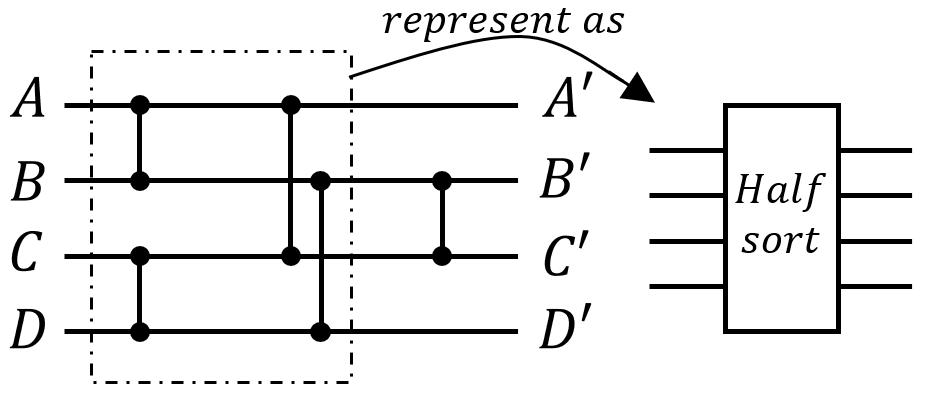}}
\caption{4-bit sorting network}
\label{fig:sortingnetwork}
\end{figure}

Fig. \ref{fig:sortingnetwork} shows a 4-input sorting network \cite{b4}. Each of the vertical lines represents a sorter in Fig. \ref{fig:sorter}. After three stages of sorting, the inputs are sorted. We know that the delay of a sorter is one stage of basic logical gate, so the sorting network consumes three stages of basic logical gates.

\subsection{Fast Carry-Generation}

In a basic full adder, suppose that the input bits are A, B, and C, the carry bit is generated by equation (\ref{equ:0}). This will consume three stages of basic logic. However, carry generation and propagation are the bottleneck of the performance in a multiplier. $AB$ represents the AND of $A$ and $B$. $A+B$ represents the OR of them.

\begin{equation}\label{equ:0}
Carry=AB + AC + BC 
\end{equation}

As we can see, the last stage in Fig. \ref{fig:sortingnetwork} only sort the second and the third bits. So it is clear that the first bit is the largest one and the fourth bit is the smallest one after two stages of sorting. Then if we choose one number randomly from the second and the third bits, this one is no lager than the first bit and no less than the fourth bit. That means the first bit, the bit randomly selected and the fourth bit have been in order, and we named them as X, Y and Z. The first two stages in Fig. \ref{fig:sorter} is represented as Half Sorter as shown in Fig. \ref{fig:sortingnetwork}. Add X, Y and Z up as binary numbers to generate carry and sum, the result is shown in Table \ref{tab:carrygeneration}. Note that X, Y, Z are in order. 

\begin{table}[htbp]
\caption{Truth table of X, Y, Z.}
\begin{center}
\begin{tabular}{|c|c|c|c|c|}
\hline
\textbf{X} & \textbf{Y} & \textbf{Z} & \textbf{Carry} & \textbf{Sum} \\
\hline
0 & 0 & 0 & 0 & 0 \\
\hline
1 & 0 & 0 & 0 & 1 \\
\hline
1 & 1 & 0 & 1 & 0 \\
\hline
1 & 1 & 1 & 1 & 1 \\
\hline
\end{tabular}
\end{center}
\label{tab:carrygeneration}
\end{table}

Because they are ordered, so there are only four possible combinations. Based on the truth table in Table \ref{tab:carrygeneration}, the bool expression is simplified as equation (\ref{equ:1}) and (\ref{equ:2}). With this special full adder structure (SFA), carry bit can be generated with two basic logic stages, faster than the traditional full adder logic which will comsume 3 logic stages.

\begin{equation}\label{equ:1}
Carry=Y 
\end{equation}

\begin{equation}\label{equ:2}
Sum=X(\overline{Y}+Z)
\end{equation}

\subsection{Adjusted Full Adder}

Then let us talk about full adders. Usually, full adder is implemented by (\ref{equ:3}) and (\ref{equ:4}). The symbol $\oplus$ means XOR. A, B and C are inputs, Carry and Sum are outputs. Sum is on the critical path. Formula (\ref{equ:4}) can be changed to: $Sum = C\oplus(A\overline{A}+A\overline{B}+\overline{A}B+B\overline{B})$. Note that $A\overline{A}+A\overline{B}+\overline{A}B+B\overline{B}=(A+B)(\overline{A}+\overline{B})=(A+B)\overline{(AB)}$. Suppose that $h_1=A+B$ and $h_2=AB$, then the formula (\ref{equ:4}) can be rewritten to formula (\ref{equ:5}). Rewrite the formula (\ref{equ:4}) to formula (\ref{equ:5}) does not reduce the logical stages, but formula (\ref{equ:5}) is more convenient with the subsequent analysis.

\begin{equation}\label{equ:3}
Carry = C(A+B)+AB 
\end{equation}

\begin{equation}\label{equ:4}
Sum = A \oplus B \oplus C 
\end{equation}

\begin{equation}\label{equ:5}
Sum=C\overline{(h_1\overline{h_2})}+\overline{C}(h_1\overline{h_2}) 
\end{equation}

\begin{figure}[htbp]
\centerline{\includegraphics[scale=0.3]{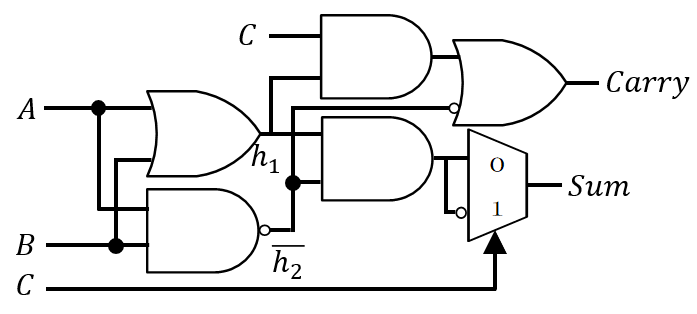}}
\caption{Adjusted full adder lgic}
\label{fig:fulladder}
\end{figure}

Fig. \ref{fig:fulladder} is the logical implementation for formula (\ref{equ:3}) and (\ref{equ:5}). As can be seen in Fig 3, if the MUX requires 2 logical stages, the sum comsumes 4 logical stages. The input signal C is only used by the MUX, and there are two logical stages from C to Sum, two logical stages from C to Carry. That means C is used after A and B, and it does not matter that if C is late to 2 logical stages of delay. The rest of this chapter will discuss how to use this feature to optimize latency.

\subsection{Implementation of (7,2) Compressor}

\begin{figure}[htbp]
\centerline{\includegraphics[scale=0.3]{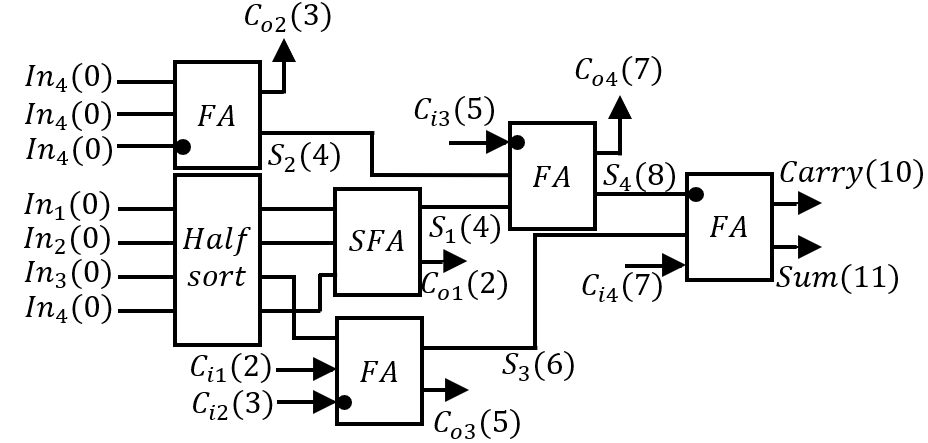}}
\caption{Overall design of (7,2) compressor}
\label{fig:overall}
\end{figure}

Fig. \ref{fig:overall} shows the overall design of (7,2) compressor. Numbers in parentheses means how many logical stages are used from input. Because of the special logic of full adder discusseed above which can generate a Carry bit with only 2 logical stages, the input signal $C_{i2}$ can be input to the full adder. This helps the whole design reduced a logical stage. Each full adder in Fig. \ref{fig:overall} have a circular marker which means the "$C$" input in Fig. \ref{fig:fulladder}. This input bit can be later than $A$ and $B$ for 2 logical stages.

By this structure, the latency of a (7,2) compressor is reduced to 11 logical stages's delay.

\begin{table*}[htbp]
\caption{Performance comparison}
\begin{center}
\begin{tabular}{|c|c|c|c|c|c|c|}
\hline
\textbf{Method} & \multicolumn{2}{|c|}{\textbf{TSMC 90nm}} & \multicolumn{2}{|c|}{\textbf{TSMC 65nm}} & \multicolumn{2}{|c|}{\textbf{TSMC 28nm}} \\
\cline{2-7} 
~ & \textbf{\textit{Delay($ps$)}}& \textbf{\textit{Area($um^2$)}}& \textbf{\textit{Delay($ps$)}}& \textbf{\textit{Area($um^2$)}}& \textbf{\textit{Delay($ps$)}}& \textbf{\textit{Area($um^2$)}} \\
\hline
[1] & 970 & 1975.0 & 550 & 1157.8 & 154 & 583.5 \\
\hline
[2] & 973 & 1822.6 & 562 &  956.2 & 155 & 559.6 \\
\hline
proposed & 937 & 2348.9 & 544 & 1055.5 & 150 & 748.6 \\
\hline
\end{tabular}
\end{center}
\label{tab:result}
\end{table*}

\section{Performance Comparison}

In this chapter, we will make a comparison with previous methods proposed in \cite{b1} and \cite{b2}. 
To make a clear comparison, all these designs will be utilized in a binary array of 7 rows and 8 columns. They are used to compress the binary array into 2 rows, just like the second step of a multiplier. Then a vector merge adder, which is implemented with Kogge-Stone algorithm, will plus them to one row. The only defference among them is the structure of (7,2) compressor. All the Verilog HDL codes are synthesised with Synopsys Design Compiler, with TSMC 90nm, 65nm and 28nm process, to find the minimum delay of each design. All the results are shown in Table \ref{tab:result}.
Methods in \cite{b1} and \cite{b2} consume 12 logical stages, so their delays are close. The proposed method in this paper comsumes 11 logical stages, so it has lower delay comprared to \cite{b1} or \cite{b2}.

As shown in Table \ref{tab:result}, the delay of this module reduced when the process nodes are reduced. But the implementations with method in \cite{b1} and \cite{b2} have almost the same delay whichever the process is used. The implementation with the proposed method have less delay than them, the saving delay is approximately one logical stage delay. Througn the data in table 2, we can sum up that with the reduction of process nodes, the influence of logic optimization decreases gradually. And because of the special full adder logic, which will comsume more ligical gates compared with tranditional full adder logic, is used, the area is larger than the methods in \cite{b1} and \cite{b2}. That means under this extreme high performance design condition, a small delay reduce requires a significant amount of area to be consumed as a cost.

\section{Conclusions}
In this paper, we proposed a special full adder logic. With its help, we proposed a new (7,2) compressor structure. This method reduces the logical stages of a (7,2) compressor to 11, less than the tranditional design methods. The special full adder logic runs with the consider of 4 bits not 3 bits, but get a quick carry out bit. Also therefore, consider 4 bits at the same time consumes more areas. So this method is suitable for high performance conditions.


\end{document}